\documentclass[prl,floatfix,showpacs,twocolumn,preprintnumbers,amsmath,amssymb,superscriptaddress]{revtex4}
\usepackage{graphicx,float,datetime}
\usepackage[ansinew]{inputenc}
\usepackage{color}
\usepackage{ragged2e}
\makeatletter
\renewcommand{\maketag@@@}[1]{\hbox{\m@th\small\normalfont#1}}%
\makeatother

\newdateformat{mydate}{\THEDAY\hspace{3pt}\monthname[\THEMONTH] \THEYEAR}

\begin{document}

\begin{center}
\title{Gravitomagnetic effects in quadratic gravity with a scalar field}
\date{\mydate\today}
\author{Andrew Finch\footnote{andrew.finch.12@um.edu.mt}}
\affiliation{Institute of Space Sciences and Astronomy, University of Malta, Msida, MSD 2080, Malta}
\affiliation{Department of Physics, University of Malta, Msida, MSD 2080, Malta}
\author{Jackson Levi Said\footnote{jackson.said@um.edu.mt}}
\affiliation{Institute of Space Sciences and Astronomy, University of Malta, Msida, MSD 2080, Malta}
\affiliation{Department of Physics, University of Malta, Msida, MSD 2080, Malta}

\begin{abstract}
{The two gravitomagnetic effects which influence bodies orbiting around a gravitational source are the geodetic effect and the Lense-Thirring effect. The former describes the precession angle of the axis of a spinning gyroscope while in orbit around a nonrotating gravitational source whereas the latter provides a correction for this angle in the case of a spinning source. In this paper we derive the relevant equations in quadratic gravity and relate them to their equivalents in general relativity. Starting with an investigation into Kepler's third law in quadratic gravity with a scalar field, the effects of an axisymmetric and rotating gravitational source on an orbiting body in a circular, equatorial orbit are introduced.}\end{abstract}


\maketitle

\end{center}

\section{I. Introduction}
\small
Over the last hundred years general relativity (GR) has passed many tests and, many of its predictions have been confirmed. Apart from gravitational waves which were detected last year (2015)  \cite{Abbott2016}, two important effects which have also been confirmed recently are the geodetic effect and the Lense-Thirring effect which together account for the total precession angle on a body in orbit around a rotating gravitational source \cite{Everitt2011}. That being said, there are areas where general relativity appears to fail to describe certain phenomena completely \cite{Clifton:2011jh}. An example where general relativity fails to agree with observation is in the case of galactic rotation curves. Along the radial coordinate the general relativistically predicted orbital speeds of masses orbiting around a disc galaxy disagree with observations unless dark matter is included \cite{Chemin2015}.

Alternative theories of gravity naturally enter the framework to either solve particular failings of the standard theory or to extend modified theories for areas where they fail or are lacking. In this paper we consider a particular alternative theory of gravity called quadratic gravity \cite{Yunes2011}. In quadratic gravity all quadratic invariants are added to the Einstein-Hilbert action where the coupling functions are dependent on a scalar field which is nonminimally coupled. The action for this model is given by \cite{Pani2011, Yunes2011}
\small
\begin{equation}\label{eq 1}
\begin{array}{l}
S=\frac{1}{16\pi}{\displaystyle\int}\sqrt{-g}d^{4}x[R-2\nabla_{\mu}\phi\nabla^{\mu}\phi-V(\phi)\\\\
\textcolor{white}{\hspace{0.6cm}}+f_{1}(\phi)R^{2}+f_{2}(\phi)R_{\mu\nu}R^{\mu\nu}+f_{3}(\phi)R_{\mu\nu\lambda\eta}R^{\mu\nu\lambda\eta}\\\\
\textcolor{white}{\hspace{0.6cm}}+f_4(\phi){R_{\mu\nu\lambda\eta}}^{*}R^{\mu\nu\lambda\eta}]+S_{mat}.\\\\
\end{array}
\end{equation}
In Eq.(\ref{eq 1}) $R$ is the the Ricci scalar, $R_{\mu\nu}$ is the Ricci tensor, $R_{\mu\nu\lambda\eta}$ is the Riemann tensor, $\phi$ is the nonminimally coupled scalar field, $V(\phi)$ is the scalar self potential and $S_{mat}$ is the matter action. The term ${R_{\mu\nu\lambda\eta}}^{*}$ is referred to as ``doubly dual" to the Riemann tensor and is obtained by applying a variation of the Levi-Civita symbol \cite{Misner1973} on the Riemann tensor. The coupling functions are taken up to linear order in their dependency on the scalar field $\phi$, therefore 
\begin{equation}\label{coupling_to_feild}
f_{i}(\phi)=\beta_{i}+\alpha_{i}\phi+\mathcal{O}(\phi^{2}),
\end{equation}
where $\alpha_{i}$ and $\beta_{i}$ are unknown coupling constants.

In this work we consider the external metric for a rotating axisymmetric gravitational source in its slowly rotating limit. As was derived by Pani \cite{Pani2011} following the field equations obtained by Yunes \cite{Yunes2011}, the ansatz metric for such a system is given by
\small
\begin{equation}\label{eq 2}
\begin{array}{l}
ds^{2}=-f(r,{\theta})dt^{2}+g(r,{\theta})^{-1}dr^{2}\\\\
\textcolor{white}{\hspace{0.85cm}}-2{W}(r)\sin^{2}{\theta}dtd\varphi+r^{2}{\Theta}(r,{\theta})d{\theta}^{2}\\\\
\textcolor{white}{\hspace{0.85cm}}+r^{2}\sin^{2}\theta{\Phi}(r,{\theta})d\varphi^{2},
\end{array}
\end{equation}
where the functions in this metric are taken to be \cite{Pani2011}
\small
\begin{equation}\label{eq 3}
\hspace{-0.08cm}\begin{array}{ll}
f(r,\theta)=&1-\dfrac{2M}{r}+\dfrac{\alpha_{3}^{2}}{4}\left[-\dfrac{49}{40M^{3}r}+\dfrac{1}{3Mr^{3}} \right. \\\\
& \left. +\dfrac{26}{3r^{4}}+\dfrac{22M}{5r^{5}}+\dfrac{32M^{2}}{5r^{6}}-\dfrac{80M^{3}}{3r^{7}}\right] \\\\
&+a^{2}\dfrac{2M\cos^{2}\theta}{r^{3}},
\end{array}
\end{equation}
\small
\begin{equation}\label{eq 4}
\hspace{-0.08cm}\begin{array}{ll}
g(r,\theta)=&1-\dfrac{2M}{r}+\dfrac{\alpha_{3}^{2}}{4}\left[-\dfrac{49}{40M^{3}r}+\dfrac{1}{M^{2}r^{2}}\right. \hfill \\\\
&\left.+\dfrac{1}{Mr^{3}}+\dfrac{52}{3r^{4}}+\dfrac{2M}{r^{5}}+\dfrac{16M^{2}}{5r^{6}}\right.\\\\
&\left.-\dfrac{368M^{3}}{3r^{7}}\right]+ a^{2}\dfrac{r-(r-2M)\cos^{2}\theta}{r^{3}},
\end{array}
\end{equation}
\small
\begin{equation}\label{eq 5}
\begin{array}{ll}
W(r)=&\dfrac{2aM}{r}-\dfrac{a\alpha^{2}_{3}}{4}\left[\dfrac{3}{5Mr^{3}}+\dfrac{28}{3r^{4}}+\dfrac{6M}{r^{5}}\right.\\\\
&\left. +\dfrac{48M^{2}}{5r^{6}}-\dfrac{80M^{3}}{3r^{7}}\right]-a\alpha^{2}_{4}\dfrac{5}{2}\left[\dfrac{1}{r^{4}}\right.\\\\
&\left.+\dfrac{12M}{7r^{5}}+\dfrac{27M^{2}}{10r^{6}}\right],
\end{array}
\end{equation}
\small
\begin{equation}\label{eq 6}
\begin{array}{l}
\Theta(r,\theta)=1+\dfrac{\cos^{2}\theta}{r^{2}}a^{2}, \\
\end{array}
\end{equation}
\small
\begin{equation}\label{eq 7}
\begin{array}{l}
\Phi(r,\theta)=1+\dfrac{r+2M\sin^{2}\theta}{r^{3}}a^{2},
\end{array}
\end{equation}
and the corresponding scalar field is given by
\small
\begin{equation}\label{eq 8}
\begin{array}{ll}
\phi(r,\theta)=&\alpha_{3}\left[\dfrac{1}{2Mr}+\dfrac{1}{2r^{2}}+\dfrac{2M}{3r^{3}}\right]\\\\
&+a\alpha_{4}\dfrac{5\cos\theta}{8M}\left[\dfrac{1}{r^{2}}+\dfrac{2M}{r^{3}}+\dfrac{18M^{2}}{5r^{4}}\right]\\\\
&-\alpha_{3}\dfrac{a^{2}}{2}\left[\dfrac{1}{10r^{4}}+\dfrac{1}{5Mr^{3}}+\dfrac{1}{4M^{2}r^{2}}+\dfrac{1}{4M^{3}r} \right. \\\\
&\left. +\cos^{2}\theta\left(\dfrac{48M}{5r^{5}}+\dfrac{21}{5r^{4}}+\dfrac{7}{5Mr^{3}}\right)\right],
\end{array}
\end{equation}
where $\alpha_{3}$ and $\alpha_{4}$ are the remaining coupling constants and $a$ is the angular momentum per unit mass defined by $a=J/M$.

We use geometric units throughout this paper, meaning that $G = c = 0$ where $G$ and $c$ are the Newtonian gravitational constant and the speed of light respectively. If the modified coupling constants are set to zero then the Kerr metric is reacquired from Eq.(\ref{eq 2}).

The paper is divided as follows: in Sec. II we derive the equation for Kepler's third law of motion. Following this, in Sec. III we consider an orbiting gyroscope and a nonrotating gravitational source which results in the geodetic effect. In Sec. IV we then consider a rotating source which leads to the Lense-Thirring precession velocity. Finally the results are discussed in Sec. V.

\section{II. Circular orbits and Kepler's Third Law}

Kepler's third law states that the square of the period of a particle in orbit around a gravitational source is proportional to the cube of its radial distance from the system's center of mass \cite{Misner1973}.

In this section we derive Kepler's third law in quadratic gravity for equatorial, circular orbits, that is, at $\theta=\frac{\pi}{2}$. The Lagrangian for this system, $\mathcal{L}$, is acquired through the substitution of the metric coefficients from Eq.(\ref{eq 2}). Taking an equatorial orbit such that $\theta=\pi/2$ and $\dot{\theta}=0$ \cite{Said2013}, the Lagrangian takes the form of
\small
\begin{equation}\label{eq 9}
\begin{array}{ll}
\mathcal{L}&=\frac{1}{2}g_{\mu\nu}\dfrac{\partial{x^{\mu}}}{\partial\tau}\dfrac{\partial{x^{\nu}}}{\partial\tau}\\\\
&=\tfrac{1}{2}\left[-f\left(r,\tfrac{\pi}{2}\right)\dot{t}^{2}+g\left(r,\tfrac{\pi}{2}\right)^{-1}\dot{r}^{2}\right.\\\\
&\textcolor{white}{\hspace{0.65cm}}\left.+r^{2}\Phi\left(r,\tfrac{\pi}{2}\right)\dot{\varphi}^{2}-2W\left(r\right)\dot{t}\dot{\varphi}\right],
\end{array}
\end{equation}  
where $g_{\mu\nu}$ are the coefficients of the metric, $\tau$ is the proper time and $x^{\sigma}$ represent the four-position $(t,r,\theta,\varphi)$. 

Being explicitly independent of $t$ and $\varphi$, the corresponding Euler-Lagrange equations of motion are given by \cite{Straumann2004}
\begin{equation}\label{eq 10}
\begin{array}{ll}
P_{t}&=E=\dfrac{\partial{\mathcal{L}}}{\partial\dot{t}}\\\\
&=-f\left(r,\tfrac{\pi}{2}\right)\dot{t}-W(r)\dot{\varphi},
\end{array}
\end{equation}
\begin{equation}\label{eq 11}
\begin{array}{ll}
P_{\varphi}&=L=\dfrac{\partial{\mathcal{L}}}{\partial\dot{\varphi}}\\\\
\vspace{2mm}&=r^{2}\Phi\left(r,\tfrac{\pi}{2}\right)\dot{\varphi}-W(r)\dot{t},
\end{array}
\end{equation}
where $E$ and $L$ are the energy and the angular momentum per unit mass, respectively. The third Euler-Lagrange equation is given by \cite{Said2013}\\
\begin{equation}\label{eq 12}
\dfrac{\partial\mathcal{L}}{\partial{r}}-\dfrac{\partial}{\partial\tau}\dfrac{\partial\mathcal{L}}{\partial\dot{r}}=0,
\end{equation}\\
which for this system gives
\begin{equation}\label{eq 13}
\begin{array}{ll}
\dfrac{\partial\mathcal{L}}{\partial{r}}=&\tfrac{1}{2}\left[\dfrac{\partial}{\partial{r}}( g^{-1}\dot{r}^{2})+\dfrac{\partial}{\partial{r}}(-f\dot{t}^{2}\right.\\\\
&\textcolor{white}{\hspace{1.6cm}}\left.+r^{2}\Phi\dot{\varphi}^{2}-2W\dot{t}\dot{\varphi})\right],\\\\
\end{array}
\end{equation}
and\\
\begin{equation}\label{eq 14}
\dfrac{\partial}{\partial\tau}\dfrac{\partial\mathcal{L}}{\partial\dot{r}}=\dfrac{\partial}{\partial\tau}\dfrac{\partial}{\partial\dot{r}}( g^{-1}\dot{r}^{2}).
\end{equation}

Due to the fact that circular orbits are being considered, the radial coordinate $r$ is also constant, $R$, implying that $\dot{r}=\frac{\partial{r}}{\partial\tau}=0$. This simplifies Eqs.(\ref{eq 13}) and (\ref{eq 14}) making them more workable. Partially expanding the functions $f$, $\Phi$ and $W$ up to their coupling constant, $\alpha_{i}$, terms, and differentiating the remaining parts of Eq.(\ref{eq 13}) with respect to $r$, one obtains
\begin{equation}\label{eq 15}
\begin{array}{ll}
\dfrac{\partial}{\partial{r}}(-f\dot{t}^{2})=\left[-\dfrac{2M}{r^{2}}-\bar{f}(R)\right]\dot{t}^{2},
\end{array}
\end{equation}
\begin{equation}\label{eq 16}
\begin{array}{ll}
\dfrac{\partial}{\partial{r}}(r^{2}\Phi\dot{\varphi}^{2})=\left[2r-\dfrac{2Ma^{2}}{r^{2}}\right]\dot{\varphi}^{2},
\end{array}
\end{equation}
\begin{equation}\label{eq 17}
\begin{array}{ll}
\dfrac{\partial}{\partial{r}}(-2W\dot{t}\dot{\varphi})=\left[\dfrac{4aM}{r^{2}}+\bar{W}(R)\right]\dot{t}\dot{\varphi},
\end{array}
\end{equation}\small
where
\begin{equation}\label{new_1}
\hspace{-0.1cm}\begin{array}{ll}
\bar{f}(R)=&\dfrac{1}{4} \alpha _3^2 \left(\dfrac{560 M^3}{3 R^8}+\dfrac{49}{40 M^3 R^2}-\dfrac{192 M^2}{5 R^7}\right.\\
&\left.-\dfrac{22 M}{R^6}-\dfrac{1}{M R^4}-\dfrac{104}{3 R^5}\right),
\end{array}
\end{equation}
\begin{equation}\label{new_2}
\begin{array}{ll}
\bar{W}(R)=&-5 a \alpha _4^2 \left(-\dfrac{81 M^2}{5 R^7}-\dfrac{60 M}{7 R^6}-\dfrac{4}{R^5}\right)\\ 
&-\dfrac{1}{2} a \alpha _3^2 \left(\dfrac{560 M^3}{3R^8}-\dfrac{288 M^2}{5 R^7}\right.\\
&\left.-\dfrac{30 M}{R^6}-\dfrac{9}{5 MR^4}-\dfrac{112}{3 R^5}\right).
\end{array}
\end{equation}
For brevity's sake the $R$ dependency is suppressed in the rest of the paper. Substituting into Eq.(\ref{eq 13}) with the first term removed and equating it to $0$ one gets a quadratic equation in terms of $\dot{\varphi}$,
\hspace{-4cm}\begin{equation}\label{eq 18}
\begin{array}{l}
\hspace{-0.4cm}\left(2r-\dfrac{2Ma^{2}}{r^{2}}\right)\dot{\varphi}^{2}+\left(\dfrac{4aM}{r^{2}}+\bar{W}\right)\dot{t}\dot{\varphi}\\ 
\hspace{3cm}-\left(\dfrac{2M}{r^{2}}+\bar{f}\right)\dot{t}^{2}=0.
\end{array}
\end{equation}

\small
Noting that $\dot{\varphi}$ is the angular velocity of the orbiting particle and solving for $\omega\equiv\dot{\varphi}$ in Eq.(\ref{eq 18}) while considering that $a \ll R$, Kepler's third law in quadratic gravity is obtained,
\begin{equation}\label{eq 19}\scriptsize
\hspace{-0.44cm}\begin{array}{ll}
\omega^{2}\approx&\dfrac{M}{R^3}+\dfrac{a^2 M^2}{R^6}-\dfrac{a^2 M \bar{f}}{2 R^4}+\dfrac{\bar{f}}{2 R}+\dfrac{a M \bar{W}}{R^4}+\dfrac{\bar{W}^2}{8 R^2}\\\\
&-\dfrac{a M \sqrt{\dfrac{16
M}{R}-\dfrac{8 a^2 M \bar{f}}{R^2}+8 R \bar{f}+\dfrac{8 a M \bar{W}}{R^2}+\bar{W}^2}}{2 R^4}\\\\
&-\dfrac{\bar{W} \sqrt{\dfrac{16 M}{R}-\dfrac{8 a^2 M \bar{f}}{R^2}+8 R \bar{f}+\dfrac{8
a M \bar{W}}{R^2}+\bar{W}^2}}{8 R^2}.
\end{array}
\end{equation}
\small
Taking $a=0$, Eq.(\ref{eq 19}) is approximately  
\begin{equation}\label{eq 20}
\begin{array}{ll}
\omega^{2}\approx&\dfrac{M}{R^3}+\dfrac{\bar{f}}{2 R}.
\end{array}
\end{equation}

Finally, taking all the coupling constants, $\alpha_{i}$, to be equal to $0$, an approximation to Kepler's law in general relativity \cite{Detweiler1989} is obtained
\begin{equation}\label{eq 21}
\begin{array}{ll}
\omega^{2}\approx&\dfrac{M}{R^3}.
\end{array}
\end{equation} 

The last term in Eq.(\ref{eq 20}) is thus the extra term given when considering quadratic gravity. As expected the effect diminishes with radial distance from the source. 
\section{III. The Geodetic effect}
The geodetic effect, also known as geodetic precession and de Sitter precession \cite{Rindler2006} gives the precession angle per orbit, $\alpha$, of the axis of an object rotating about a nonrotating gravitational source. In this section we calculate the angle for a circular equatorial orbit in quadratic gravity. 
\small
Consider the equatorial plane for this metric, as well as the transformation of the $\varphi$-angle for a rotating observer, that is $\varphi\rightarrow \varphi+\omega t$, which gives 
\begin{equation}\label{eq 23}
\begin{array}{ll}
ds^{2}&=-f\left(r,\tfrac{\pi}{2}\right)dt^{2}+g\left(r,\tfrac{\pi}{2}\right)^{-1}dr^{2}\\\\
&\textcolor{white}{\hspace{3cm}}+r^{2}(d\varphi+\omega{dt})^2, 
\end{array}
\end{equation}
where $\omega$ is the angular velocity, and $f$ and $g$ have been evaluated at the equator. By expanding and using the difference of two squares method on Eq.(\ref{eq 23}) a different form for the metric is obtained
\begin{equation}\label{eq 24}
\begin{array}{ll}
ds^{2}=&-(f-r^{2}\omega^{2})\left[dt-\dfrac{r^{2}\omega}{f-r^{2}\omega^{2}}d\varphi\right]^{2}\\\\
&+r^{2}\left[\dfrac{f}{f-r^{2}\omega^{2}}\right]d\varphi^{2}+g^{-1}dr^{2}.
\end{array}
\end{equation}
This form can be compared to Rindler's canonical form \cite{Rindler2006}
\begin{equation}\label{eq 25}
\begin{array}{ll}
ds^{2}=-e^{2\lambda}(dt-\text{w}_{i}dx^{i})^{2}+k_{ij}dx^{i}dx^{j},
\end{array}
\end{equation}
where the latin indices refer to the spacial coordinates only, that is, $r$, $\theta$ and $\varphi$. When comparing Eq.(\ref{eq 24}) to Eq.(\ref{eq 25}) the following coefficients for the nonvanishing coordinates are obtained
\begin{equation}\label{eq 26}
e^{2\lambda}=f-r^{2}\omega^{2},
\end{equation}
\begin{equation}\label{eq 27}
\text{w}_{3}=\dfrac{r^{2}\omega}{f-r^{2}\omega^{2}},
\end{equation}
\begin{equation}\label{eq 28}
k_{11}=g^{-1},
\end{equation}
\begin{equation}\label{eq 29}
k_{33}=\dfrac{r^{2}f}{f-r^{2}\omega^{2}}.
\end{equation}

Since free circular orbits are being considered the object experiences no change in motion resulting in a vanishing acceleration \cite{Rindler2006,Said2013}, that is
\begin{equation}\label{eq 30}
a=(k^{ij}\lambda_{,i}\lambda_{,j})^{\tfrac{1}{2}}=0.
\end{equation}
Raising the indices of $k_{ij}$ gives 
\begin{equation}\label{eq 31}
\begin{array}{ll}
k^{11}=&g^{1\mu}g^{\nu{1}}k_{11}=g,
\end{array}
\end{equation}
\begin{equation}\label{eq 32}
\begin{array}{ll}
k^{33}=&g^{3\mu}g^{\nu{3}}k_{33}=\dfrac{f-r^{2}\omega^{2}}{r^{2}f},
\end{array}
\end{equation}
and determining which derivatives do not vanish, one obtains
\begin{equation}\label{eq 33}
\begin{array}{ll}
a&=(k^{11}\lambda_{,1}\lambda_{,1})^{1/2}+(k^{33}\lambda_{,3}\lambda_{,3})^{1/2}\\
&=(k^{11}\lambda_{,1}\lambda_{,1})^{1/2}\\
&=0,
\end{array}
\end{equation}
since $\lambda_{,3}=0$ as can be seen from Eq.(\ref{eq 26}). From Eq.(\ref{eq 31}), $k^{11}$ is clearly not equal to $0$ and thus $\lambda_{,1}$ must be. From Eq.(\ref{eq 26}) it follows that
\begin{equation}\label{eq 34}
\lambda=\tfrac{1}{2}\ln(f-r^{2}\omega^{2}),
\end{equation}
Substituting Eqs.(\ref{eq 31}) and (\ref{eq 34}) into Eq.(\ref{eq 33}) results in
\begin{equation}\label{eq 35}
\begin{array}{ll}
\dfrac{\partial}{\partial{r}}(f-r^{2}\omega^{2})&=\dfrac{\partial{f}}{\partial{r}}-2r\omega^{2}\\
&=F-2r\omega^{2}\\
&=0.
\end{array}
\end{equation}

Rearranging, this gives an equation for the square of the angular velocity $\omega$ which is comparable with the one obtained in the derivation for Kepler's third law in quadratic gravity with $a=0$,
\begin{equation}\label{eq 36}
\omega^{2}=\dfrac{F}{2r}.
\end{equation}
Substituting Eq.(\ref{eq 36}) into Eqs.(\ref{eq 26}) and (\ref{eq 32}) gives
\begin{equation}\label{eq 37}
e^{2\lambda}=f-\dfrac{rF}{2},
\end{equation}
\begin{equation}\label{eq 38}
k^{33}=\dfrac{f-\frac{rF}{2}}{r^{2}f}.
\end{equation}

The general equation for the rotation rate of a gyrocompass in proper time as given by Rindler is \cite{Rindler2006}
\begin{equation}\label{eq 39}
\Omega=\tfrac{1}{2\sqrt{2}}e^{\lambda}[k^{im}k^{jl}(\text{w}_{i,j}-\text{w}_{j,i})(\text{w}_{m,l}-\text{w}_{l,m})]^{1/2},
\end{equation}
which in this case reduces to
\begin{equation}\label{eq 40}
\Omega=\tfrac{e^{\lambda}}{2}[k^{11}k^{33}{\text{w}_{3,1}}^2]^{1/2}.
\end{equation}
Differentiating $\text{w}_{3}$ and substituting $e^{\lambda}$, $k^{11}$ and $k^{33}$ in Eq.(\ref{eq 40}) gives
\begin{equation}\label{eq 41}
\Omega=\omega\left(\dfrac{g}{f}\right)^{1/2}.
\end{equation}

The square of the coordinate angular velocity $\omega$ as determined in Eq.(\ref{eq 36}) relates the coordinate time $t$ and the angular coordinate $\varphi$ through
\begin{equation}\label{eq 42}
\begin{array}{rlll}
&\omega^{2}&=\left(\dfrac{\partial{\varphi}}{\partial{t}}\right)^{2}=&\dfrac{F}{2r},\\\\
\Rightarrow &d\varphi^{2}&=\dfrac{F}{2R} dt^{2}.&
\end{array}
\end{equation}

Consider circular equatorial orbits, meaning that $dr = 0$ and $d\theta = 0$, respectively. Substituting $d\varphi^2$ in the original metric, Eq.(\ref{eq 2}), gives
\begin{equation}\label{eq 43}
\begin{array}{ll}
ds^{2}&=-fdt^{2}+\dfrac{R^{2}F}{2R}dt^{2}\\
&=\left(\dfrac{RF}{2}-f\right)dt^{2},
\end{array}
\end{equation}
\vspace{-1mm}\begin{equation}\label{eq 44}
\Delta\tau=\left(f-\dfrac{RF}{2}\right)^{1/2}\Delta{t}.
\end{equation}

Using Eqs.(\ref{eq 41}) and (\ref{eq 44}) the precession angle for one complete orbit with respect to the rotating frame is found to be
\begin{equation}\label{eq 45}
\begin{array}{ll}
\alpha'&=\Omega\Delta\tau\\
&=\omega\left(\dfrac{g}{f}\right)^{1/2}\left(f-\dfrac{RF}{2}\right)^{1/2}\dfrac{2\pi}{\omega}\\
&\approx 2\pi\left(\dfrac{g}{f}\right)^{1/2}\left[1+\dfrac{(f-1)}{2}-\dfrac{RF}{4}\right].\\\\\\
\end{array}
\end{equation}
As a result the precession angle per orbit, and thus the angle required, is
\begin{equation}\label{eq 46}
\begin{array}{ll}
\alpha&=2\pi-\alpha'\\
&=2\pi\left(\dfrac{g}{f}\right)^{1/2}\left[\dfrac{(1-f)}{2}+\dfrac{RF}{4}\right].\\
\end{array}
\end{equation}

The only quadratic gravity parameter left in this equation is $\alpha_{3}$. When taking $\alpha_{3}=0$ Eq.(\ref{eq 46}) reduces to 
\begin{equation}\label{eq 47}
\alpha=\dfrac{3M\pi}{R},
\end{equation}
which is the general relativistic result for the geodetic effect as described by Rindler in Ref.\cite{Rindler2006}, as expected.

\vspace{0.1cm} 

\section{IV. Lense-Thirring Effect}

The Lense-Thirring effect, also known as the frame dragging effect, is a correction for the precession of a spinning gyroscope orbiting around a rotating source \cite{Schutz2009}. As in \cite{Tartaglia1998,Said2013,Nandi:2000xt} we consider the Sagnac effect in order to calculate the Lense-Thirring precession velocity. Through the Sagnac effect we consider two rotating light beams emitted simultaneously from the same source which are then allowed to corotate and counter-rotate respectively. The source can also act as a receiver. 

We again consider a circular equatorial orbit but this time with a rotating gravitational source, that is $a \neq 0$, thus giving
\begin{equation}\label{eq 48}
\begin{array}{l}
\hspace{-0.1cm}ds^{2}=-f\left(R,\tfrac{\pi}{2}\right)dt^{2}-2W(R)dtd\varphi\\\\
\textcolor{white}{\hspace{3.6cm}}+R^{2}\Phi\left(R,\tfrac{\pi}{2}\right)d\varphi^{2}.\\
\end{array}
\end{equation}

Assuming that the rotation of the light source is uniform, the rotation angle of the light source/receiver, $\varphi_{o}$, can be described by $\varphi_{o}=\omega_{o}t$. On taking the derivative we find $d\varphi_{o}=\omega_{o}dt$, which when substituting into Eq.(\ref{eq 48}) results in
\begin{equation}\label{eq 49}
\begin{array}{l}
\hspace{-0.1cm}ds^{2}=-f\left(R,\tfrac{\pi}{2}\right)dt^{2}-2\omega_{o}W(R)dt^{2}\\\\
\textcolor{white}{\hspace{3.2cm}}+R^{2}{\omega_{o}}^{2}\Phi\left(R,\tfrac{\pi}{2}\right)dt^{2}.\\
\end{array}
\end{equation}

Given that these are light rays, $ds=0$, Eq.(\ref{eq 49}) turns into a quadratic equation in $\omega_{o}$. Solving for $\omega_{o}$ one obtains the following equation for the angular velocity of the light beams,
\vspace{0.01cm}
\begin{widetext}
\begin{equation}\label{eq 50}
\begin{array}{ll}
\Omega_{\pm}=&\dfrac{1}{a^2+\dfrac{2 a^2 M}{R}+R^2}\left\{\left[\dfrac{2 a M}{R}+A \alpha _3^2+B \alpha _4^2\right]\right.\\\\
&\textcolor{white}{\hspace{3cm}}\left.\pm \sqrt{\left(a^2-2 M R+R^2\right)+C \alpha _3^2+D \alpha _3^4+H \alpha _4^2+F \alpha
_3^2 \alpha _4^2+G \alpha _4^4}\right\},
\end{array}
\end{equation}
\newpage
\end{widetext}

\noindent where $A(R,M,a)$ through to $G(R,M,a)$ have suppressed arguments for brevity's sake. Their exact dependencies are given by 

\begin{equation}\label{B3}
\begin{array}{ll}
\hspace{-0.363cm}A(R,M,a)=& \dfrac{20 a M^3}{3 R^7}-\dfrac{12 a M^2}{5 R^6}-\dfrac{3 a M}{2 R^5}\\\\
&-\dfrac{7 a}{3 R^4}-\dfrac{3 a}{20 M R^3},\\\\
\end{array}
\end{equation}
\begin{equation}\label{B4}
\hspace{1mm}\begin{array}{ll}
B(R,M,a)=&-\dfrac{27 a M^2}{4 R^6}-\dfrac{30 a M}{7 R^5}-\dfrac{5 a}{2 R^4},\\\\
\end{array}
\end{equation}
\begin{equation}\label{B5}
\hspace{-3mm}\begin{array}{ll}
C(R,M,a)=& \dfrac{40 a^2 M^4}{3 R^8}-\dfrac{196 a^2 M^3}{15 R^7}\\\\
&-\dfrac{11 a^2 M^2}{5 R^6}-\dfrac{39 a^2 M}{10 R^5}-\dfrac{20 M^3}{3
R^5}\\\\
&+\dfrac{26 a^2}{15 R^4}+\dfrac{8 M^2}{5 R^4}+\dfrac{a^2}{12 M R^3}\\\\
&+\dfrac{11 M}{10 R^3}+\dfrac{13}{6 R^2}-\dfrac{49 a^2}{80 M^2 R^2}\\\\
&-\dfrac{49 a^2}{160 M^3 R}+\dfrac{1}{12 M R}-\dfrac{49 R}{160 M^3},\\\\
\end{array}
\end{equation}

\begin{equation}\label{B6}
\hspace{0cm}\begin{array}{ll}
D(R,M,a)=& \dfrac{400 a^2 M^6}{9 R^{14}}-\dfrac{32 a^2 M^5}{R^{13}}-\dfrac{356 a^2 M^4}{25 R^{12}}\\\\
&-\dfrac{1076 a^2 M^3}{45 R^{11}}+\dfrac{229
a^2 M^2}{20 R^{10}}\\\\
&+\dfrac{193 a^2 M}{25 R^9}+\dfrac{1061 a^2}{180 R^8}+\dfrac{7 a^2}{10 M R^7}\\\\
&+\dfrac{9 a^2}{400 M^2 R^6},\\\\
\end{array}
\end{equation}
\begin{equation}\label{B7}
\hspace{-0.95cm}\begin{array}{ll}
H(R,M,a)=&-\dfrac{27 a^2 M^3}{R^7}-\dfrac{120 a^2 M^2}{7 R^6}\\\\
&-\dfrac{10 a^2 M}{R^5},\\\\
\end{array}
\end{equation}
\begin{equation}\label{B8}
\begin{array}{ll}
F(R,M,a)=& -\dfrac{90 a^2 M^5}{R^{13}}-\dfrac{866 a^2 M^4}{35 R^{12}}\\\\
&+\dfrac{629 a^2 M^3}{84 R^{11}}+\dfrac{789 a^2 M^2}{14 R^{10}}\\\\
&+\dfrac{1181 a^2 M}{40 R^9}+\dfrac{272 a^2}{21 R^8}+\dfrac{3 a^2}{4 M R^7},\\\\
\end{array}
\end{equation}
\begin{equation}\label{B9}
\begin{array}{ll}
G(R,M,a)=& \dfrac{729 a^2 M^4}{16 R^{12}}+\dfrac{405 a^2 M^3}{7 R^{11}}\\\\
&+\dfrac{10215 a^2 M^2}{196 R^{10}}+\dfrac{150 a^2 M}{7 R^9}\\\\
&+\dfrac{25a^2}{4 R^8},\\\\
\end{array}
\end{equation}

From Eq.(\ref{eq 50}) the rotation angle for the two light rays is found to be $\varphi_{\pm}=\Omega_{\pm}t$. Combining this with the equation for the rotation angle of the light source/receiver, $\varphi_{o}=\omega_{o}t$, results in Eq.(\ref{eq 51})
\begin{equation}\label{eq 51}
\varphi_{\pm}=\dfrac{\Omega_{\pm}}{\omega_{o}}\varphi_{o},
\end{equation}
where the $\pm$ signs represent the corotating and counter-rotating light ray directions, respectively. The first time the two rays pass through the receiver is when the $\phi$ angles assume the respective values
\begin{equation}\label{eq 52}
\begin{array}{ll}
\varphi_{+}=&\varphi_{o}+2\pi,\\
\varphi_{-}=&\varphi_{o}-2\pi.
\end{array}
\end{equation}
Equating Eq.(\ref{eq 51}) and Eq.(\ref{eq 52}), and rearranging the result in order to obtain an equation for $\varphi_{o\pm}$ gives
\begin{equation}\label{eq 53} 
\varphi_{o\pm}=\dfrac{\pm{2}\pi\omega_{o}}{\Omega_{\pm}-\omega_{o}}.
\end{equation}
Substituting for $\Omega_{\pm}$ this reduces to

\begin{widetext}

\footnotesize

\begin{equation}\label{eq 54} 
\hspace{-0.8cm}\begin{array}{l}
\textcolor{white}{\hspace{0.65cm}}\varphi_{o\pm}=\dfrac{\pm{2}\pi\omega_{o}}{\dfrac{1}{a^2+\dfrac{2 a^2 M}{R}+R^2}\left\{\left[\dfrac{2 a M}{R}+A \alpha _3^2+B \alpha _4^2\right]\pm \sqrt{\left(a^2-2 M R+R^2\right)+C \alpha _3^2+D \alpha _3^4+H \alpha _4^2+F \alpha_3^2 \alpha _4^2+G \alpha _4^4}\right\}-\omega_{o}}.\\\\
\end{array}
\end{equation}

\end{widetext}

\small

Noting that $ds^{2}=-d\tau^{2}$ and $d\varphi=\omega_{o}dt$, the metric in Eq.(\ref{eq 49}) takes on the form
\begin{equation}\label{eq 55} 
d\tau=\sqrt{\left({f+\omega_{o}2W-\omega_{o}}^{2}R^{2}\Phi\right)} \;\; \dfrac{d\varphi_{o}}{\omega_{o}}.
\end{equation}
Integrating from $\varphi_{o-}$ to $\varphi_{o+}$ gives
\begin{equation}\label{eq 56}
\delta\tau=\sqrt{\left({f+\omega_{o}2W-\omega_{o}}^{2}R^{2}\Phi\right)} \;\; \dfrac{\varphi_{o+} - \varphi_{o-} }{\omega_{o}},
\end{equation}
which represents the time delay between the arrival times of the light beams. Substituting for $\varphi_{o+}$ and $\varphi_{o-}$ and rearranging gives the final form of the time delay. This gives

\begin{widetext}

\begin{equation}\label{eq 57}
\hspace{-1cm}\begin{array}{ll}
\delta \tau &=\left(\dfrac{4 \pi }{R}\right)\;\;\dfrac{ \left[\omega_{o} \left(R a^2+2 a^2 M+R^3\right)-2 a M-R \left(\alpha _3^2 A+\alpha _4^2 B\right)\right]}{\sqrt{{f+\omega_{o}2W-\omega_{o}}^{2}R^{2}\Phi}}.
\end{array}
\end{equation}

\end{widetext}

It can then be deduced that when the angular velocity of the source/receiver is
\begin{equation}\label{eq 59}
\omega_{o}=\dfrac{2 a M+R \left(\alpha _3^2 A+\alpha _4^2 B\right)}{R a^2+2 a^2 M+R^3},
\end{equation}
$\delta\tau$ is $0$. This case is equivalent to the case where an observer is in an orbit with the same angular velocity as the gravitational source and thus no difference in arrival time is recorded. Dividing the numerator and denominator of Eq.(\ref{eq 57}) by $R^{3}$ and taking the quadratic gravity coupling constants $\alpha_{3}$ and $\alpha_{4}$ to be equal to zero gives the same $\omega$ as obtained in by Tartaglia in Ref.\cite{Tartaglia1998} recovering GR.

Considering any other angular velocity and taking $a\neq0$, $\delta\tau$ does not reduce to zero. As previously deduced, $\omega_0$ can be treated as the angular velocity of the observer around the source and thus for an observer at a fixed position with respect to distant stars $\omega_{0}=0$ results in
\begin{equation}\label{eq 60}
\begin{array}{ll}
\delta\tau_{w=0}=\delta\tau_{o}&=-\left(\dfrac{4\pi}{R}\right)\dfrac{2aM+R(\alpha _3^2 A+\alpha _4^2 B)}{\sqrt{f\left(r,\tfrac{\pi}{2}\right)}}\\\\
&=-\dfrac{8\pi}{R}\dfrac{Ma}{\sqrt{f}}-\dfrac{4\pi(\alpha _3^2 A+\alpha _4^2 B)}{\sqrt{f}}\\\\
&=-\dfrac{8\pi}{R}\dfrac{J}{\sqrt{f}}-\dfrac{4\pi(\alpha _3^2 A+\alpha _4^2 B)}{\sqrt{f}}.
\end{array}
\end{equation}

This represents the time delay effect due solely to the rotation of the gravitational source, i.e. the frame dragging by the $a\neq0$ parameter and not the light ray in orbit about this body. Hence this time delay is the quantity we are interested in in this section.

Finally the difference in arrival time for the light rays due to the rotation of the source alone can be expressed in terms of the Lense-Thirring precession velocity, which is given by
\begin{equation}\label{eq 61}
\omega_{LT}=-\dfrac{J+\tfrac{1}{2}R(\alpha _3^2 A+\alpha _4^2 B)}{R^3}.
\end{equation}
As in Ref.\cite{Said2013} $\omega_{LT}$ is determined on comparison with the GR case. Substituting in the time delay formula
\begin{equation}\label{eq 62}
\delta\tau_{o}={8\omega_{LT}}\dfrac{\pi{R^{2}}}{\sqrt{f}}.
\end{equation}
Consequently as $\alpha_{i}$ vanishes $\omega_{LT}$ reduces to
\begin{equation}\label{eq 63}
\omega_{LT}=-\dfrac{J}{R^3},
\end{equation}
while the Sagnac time delay turns out to be
\begin{equation}\label{eq 64}
\delta\tau_{o}=8\omega_{LT}\dfrac{\pi{R^{2}}}{\sqrt{1-\dfrac{2M}{R}}},
\end{equation}
which both agree with the situation in GR \cite{Tartaglia1998}.

\section{V. Discussion and conclusion}

In this paper we have considered three orbital effects in quadratic gravity, namely, Kepler's third law and the two gravitomagnetic effects, namely the geodetic precession and the Lense-Thirring effect, which can be found in Eqs.(\ref{eq 19}), (\ref{eq 46}) and (\ref{eq 61}), respectively. Starting with Kepler's third law, circular orbits are investigated with the angular velocity with respect to coordinate time being found as a function of the constant radius $R$. The result is in agreement with Kepler's law for the general relativity case.

As expected the correction keeps to our intuitive picture of how the effect works, i.e. it continues to diminish with distance. In the far field the additional terms add more significantly to the effect.

When considering the geodetic effect, we obtained Eq.(\ref{eq 46}), which was only dependent on one of the coupling constants, $\alpha_3$. This came about as a result of the derivation only involving the radial and coordinate time metric entries, on the equatorial plane. 

Lastly we considered the Lense-Thirring effect. We did this using the Sagnac effect which computes the time delay between arrival times for counter rotating beams of light, as shown in Eq.(\ref{eq 60}). This was then related to the Lense-Thirring precession velocity in Eq.(\ref{eq 61}) which is in agreement with the general relativity case represented in Ref.\cite{Tartaglia1998}. Along with the standard term ,$-J/R^3$, we also find, $\tfrac{1}{2}(\alpha _3^2 A+\alpha _4^2 B)/R^2$, which comes about only for nonvanishing $\alpha_3$ and $\alpha_4$ coupling constants. 

Currently there are no value ranges for the coupling parameters $\alpha_3$ and $\alpha_4$. Given the close correlation of the recent Gravity Probe B experiment \cite{Everitt2011} results with the GR prediction the modified terms are expected to be small. However these terms may play an important role for more exotic events. 

\newpage
\section{Acknowledgments}
A.F thanks the Institute of Space Sciences and Astronomy at the University of Malta for its support and for the internship granted during the completion of this work. This work was supported in part by UoM Grant No. SSARP01-16.


\begin{thebibliography}{14}
\expandafter\ifx\csname natexlab\endcsname\relax\def\natexlab#1{#1}\fi
\expandafter\ifx\csname bibnamefont\endcsname\relax
  \def\bibnamefont#1{#1}\fi
\expandafter\ifx\csname bibfnamefont\endcsname\relax
  \def\bibfnamefont#1{#1}\fi
\expandafter\ifx\csname citenamefont\endcsname\relax
  \def\citenamefont#1{#1}\fi
\expandafter\ifx\csname url\endcsname\relax
  \def\url#1{\texttt{#1}}\fi
\expandafter\ifx\csname urlprefix\endcsname\relax\def\urlprefix{URL }\fi
\providecommand{\bibinfo}[2]{#2}
\providecommand{\eprint}[2][]{\url{#2}}

\bibitem[{\citenamefont{Abbott et~al.}(2016)\citenamefont{Abbott, Abbott,
  Abbott, Abernathy, Acernese, Ackley, Adams, Adams, Addesso, Adhikari
  et~al.}}]{Abbott2016}
\bibinfo{author}{\bibfnamefont{B.~P.} \bibnamefont{Abbott}},
  \bibinfo{author}{\bibfnamefont{R.}~\bibnamefont{Abbott}},
  \bibinfo{author}{\bibfnamefont{T.~D.} \bibnamefont{Abbott}},
  \bibinfo{author}{\bibfnamefont{M.~R.} \bibnamefont{Abernathy}},
  \bibinfo{author}{\bibfnamefont{F.}~\bibnamefont{Acernese}},
  \bibinfo{author}{\bibfnamefont{K.}~\bibnamefont{Ackley}},
  \bibinfo{author}{\bibfnamefont{C.}~\bibnamefont{Adams}},
  \bibinfo{author}{\bibfnamefont{T.}~\bibnamefont{Adams}},
  \bibinfo{author}{\bibfnamefont{P.}~\bibnamefont{Addesso}},
  \bibinfo{author}{\bibfnamefont{R.~X.} \bibnamefont{Adhikari}},
  \bibnamefont{et~al.} (\bibinfo{collaboration}{LIGO Scientific Collaboration
  and Virgo Collaboration}), \bibinfo{journal}{Phys. Rev. Lett.}
  \textbf{\bibinfo{volume}{116}}, \bibinfo{pages}{061102}
  (\bibinfo{year}{2016}).

\bibitem[{\citenamefont{Everitt et~al.}(2011)\citenamefont{Everitt, DeBra,
  Parkinson, Turneaure, Conklin, Heifetz, Keiser, Silbergleit, Holmes,
  Kolodziejczak et~al.}}]{Everitt2011}
\bibinfo{author}{\bibfnamefont{C.~W.~F.} \bibnamefont{Everitt}},
  \bibinfo{author}{\bibfnamefont{D.~B.} \bibnamefont{DeBra}},
  \bibinfo{author}{\bibfnamefont{B.~W.} \bibnamefont{Parkinson}},
  \bibinfo{author}{\bibfnamefont{J.~P.} \bibnamefont{Turneaure}},
  \bibinfo{author}{\bibfnamefont{J.~W.} \bibnamefont{Conklin}},
  \bibinfo{author}{\bibfnamefont{M.~I.} \bibnamefont{Heifetz}},
  \bibinfo{author}{\bibfnamefont{G.~M.} \bibnamefont{Keiser}},
  \bibinfo{author}{\bibfnamefont{A.~S.} \bibnamefont{Silbergleit}},
  \bibinfo{author}{\bibfnamefont{T.}~\bibnamefont{Holmes}},
  \bibinfo{author}{\bibfnamefont{J.}~\bibnamefont{Kolodziejczak}},
  \bibnamefont{et~al.}, \bibinfo{journal}{Phys. Rev. Lett.}
  \textbf{\bibinfo{volume}{106}}, \bibinfo{pages}{221101}
  (\bibinfo{year}{2011}).

\bibitem[{\citenamefont{Clifton et~al.}(2012)\citenamefont{Clifton, Ferreira,
  Padilla, and Skordis}}]{Clifton:2011jh}
\bibinfo{author}{\bibfnamefont{T.}~\bibnamefont{Clifton}},
  \bibinfo{author}{\bibfnamefont{P.~G.} \bibnamefont{Ferreira}},
  \bibinfo{author}{\bibfnamefont{A.}~\bibnamefont{Padilla}}, \bibnamefont{and}
  \bibinfo{author}{\bibfnamefont{C.}~\bibnamefont{Skordis}},
  \bibinfo{journal}{Phys. Rep.} \textbf{\bibinfo{volume}{513}},
  \bibinfo{pages}{1} (\bibinfo{year}{2012}), \eprint{1106.2476}.

\bibitem[{\citenamefont{{Chemin} et~al.}(2015)\citenamefont{{Chemin}, {Renaud},
  and {Soubiran}}}]{Chemin2015}
\bibinfo{author}{\bibfnamefont{L.}~\bibnamefont{{Chemin}}},
  \bibinfo{author}{\bibfnamefont{F.}~\bibnamefont{{Renaud}}}, \bibnamefont{and}
  \bibinfo{author}{\bibfnamefont{C.}~\bibnamefont{{Soubiran}}},
  \bibinfo{journal}{Astron. Astrophys.} \textbf{\bibinfo{volume}{578}},
  \bibinfo{eid}{A14} (\bibinfo{year}{2015}), \eprint{1504.01507}.

\bibitem[{\citenamefont{Yunes and Stein}(2011)}]{Yunes2011}
\bibinfo{author}{\bibfnamefont{N.}~\bibnamefont{Yunes}} \bibnamefont{and}
  \bibinfo{author}{\bibfnamefont{L.~C.} \bibnamefont{Stein}},
  \bibinfo{journal}{Phys. Rev. D} \textbf{\bibinfo{volume}{83}},
  \bibinfo{pages}{104002} (\bibinfo{year}{2011}).

\bibitem[{\citenamefont{Pani et~al.}(2011)\citenamefont{Pani, Macedo, Crispino,
  and Cardoso}}]{Pani2011}
\bibinfo{author}{\bibfnamefont{P.}~\bibnamefont{Pani}},
  \bibinfo{author}{\bibfnamefont{C.~F.~B.} \bibnamefont{Macedo}},
  \bibinfo{author}{\bibfnamefont{L.~C.~B.} \bibnamefont{Crispino}},
  \bibnamefont{and} \bibinfo{author}{\bibfnamefont{V.}~\bibnamefont{Cardoso}},
  \bibinfo{journal}{Phys. Rev. D} \textbf{\bibinfo{volume}{84}},
  \bibinfo{pages}{087501} (\bibinfo{year}{2011}).

\bibitem[{\citenamefont{Misner et~al.}(1973)\citenamefont{Misner, Thorne, and
  Wheeler}}]{Misner1973}
\bibinfo{author}{\bibfnamefont{C.}~\bibnamefont{Misner}},
  \bibinfo{author}{\bibfnamefont{K.}~\bibnamefont{Thorne}}, \bibnamefont{and}
  \bibinfo{author}{\bibfnamefont{J.}~\bibnamefont{Wheeler}}, Gravitation
  (\bibinfo{publisher}{W. H. Freeman, New York}, \bibinfo{year}{1973}).

\bibitem[{\citenamefont{Said et~al.}(2013)\citenamefont{Said, Sultana, and
  Adami}}]{Said2013}
\bibinfo{author}{\bibfnamefont{J.~L.} \bibnamefont{Said}},
  \bibinfo{author}{\bibfnamefont{J.}~\bibnamefont{Sultana}}, \bibnamefont{and}
  \bibinfo{author}{\bibfnamefont{K.~Z.} \bibnamefont{Adami}},
  \bibinfo{journal}{Phys. Rev. D} \textbf{\bibinfo{volume}{88}},
  \bibinfo{pages}{087504} (\bibinfo{year}{2013}).

\bibitem[{\citenamefont{Straumann}(2013)}]{Straumann2004}
\bibinfo{author}{\bibfnamefont{N.}~\bibnamefont{Straumann}},
  \emph{\bibinfo{title}{General Relativity: With Applications to
  Astrophysics}}, Theoretical and Mathematical Physics
  (\bibinfo{publisher}{Springer Berlin Heidelberg}, \bibinfo{year}{2013}).

\bibitem[{\citenamefont{Evans et~al.}(1989)\citenamefont{Evans, Finn, and
  Hobill}}]{Detweiler1989}
\bibinfo{author}{\bibfnamefont{C.}~\bibnamefont{Evans}},
  \bibinfo{author}{\bibfnamefont{L.}~\bibnamefont{Finn}}, \bibnamefont{and}
  \bibinfo{author}{\bibfnamefont{D.}~\bibnamefont{Hobill}},
  \emph{\bibinfo{title}{Frontiers in Numerical Relativity}}
  (\bibinfo{publisher}{Cambridge University Press, Cambridge},
  \bibinfo{year}{1989}).

\bibitem[{\citenamefont{Rindler}(2006)}]{Rindler2006}
\bibinfo{author}{\bibfnamefont{W.}~\bibnamefont{Rindler}},
  \emph{\bibinfo{title}{Relativity: Special, General, and Cosmological}}
  (\bibinfo{publisher}{Oxford University Press, Oxford}, \bibinfo{year}{2006}).

\bibitem[{\citenamefont{Schutz}(2009)}]{Schutz2009}
\bibinfo{author}{\bibfnamefont{B.}~\bibnamefont{Schutz}},
  \emph{\bibinfo{title}{A First Course in General Relativity}}
  (\bibinfo{publisher}{Cambridge University Press, Cambridge},
  \bibinfo{year}{2009}).

\bibitem[{\citenamefont{Tartaglia}(1998)}]{Tartaglia1998}
\bibinfo{author}{\bibfnamefont{A.}~\bibnamefont{Tartaglia}},
  \bibinfo{journal}{Phys. Rev. D} \textbf{\bibinfo{volume}{58}},
  \bibinfo{pages}{064009} (\bibinfo{year}{1998}).

\bibitem[{\citenamefont{Nandi et~al.}(2001)\citenamefont{Nandi, Alsing, Evans,
  and Nayak}}]{Nandi:2000xt}
\bibinfo{author}{\bibfnamefont{K.~K.} \bibnamefont{Nandi}},
  \bibinfo{author}{\bibfnamefont{P.~M.} \bibnamefont{Alsing}},
  \bibinfo{author}{\bibfnamefont{J.~C.} \bibnamefont{Evans}}, \bibnamefont{and}
  \bibinfo{author}{\bibfnamefont{T.~B.} \bibnamefont{Nayak}},
  \bibinfo{journal}{Phys. Rev.} \textbf{\bibinfo{volume}{D 63}},
  \bibinfo{pages}{084027} (\bibinfo{year}{2001}).

\end{thebibliography}
\end{document}